# Bolometric Superconducting Optical Nanoscopy (BOSON)

Ran Jing[1,2†], Boyi Zhou[1,3†], Dingchen Kang[1], Wenjun Zheng[1], Zijian Zhou[1], Heng Wang[1], Xinzhong Chen[1], Juntao Yao[2,4], Bing Cheng[1], Ji-Hoon Park[1], Lukas Wehmeier[6], Zhenbing Dai[1], Shoujing Chen[1], Christopher D. Prainito[1], G. L. Carr[6], Ilya Charaev[7], Denis Bandurin[8], Genda Gu[2], Qiang Li[1,2], Karl. K. Berggren[9], D. N. Basov[5*], Xu Du[1], Mengkun Liu[1,6*]

[1] Department of Physics and Astronomy, Stony Brook University; Stony Brook, New York 11794, USA

[2] Condensed Matter Physics and Materials Science Division, Brookhaven National Laboratory; Upton, NY, 11973-5000, USA.

[3] Center for Integrated Science and Engineering, Columbia University, New York, New York 10027, USA

[4] Department of Materials Science and Chemical Engineering, Stony Brook University, Stony Brook, New York, 11794 USA

[5] Department of Physics, Columbia University; New York, New York 10027, USA.

[6] National Synchrotron Light Source II, Brookhaven National Laboratory; Upton, New York 11973, USA.

[7] Physics Institute, University of Zurich, Zurich 8057, Switzerland

[8] Department of Materials Science and Engineering, National University of Singapore, 117551, Singapore

[9] Department of Electrical Engineering, Massachusetts Institute of Technology, Cambridge, MA 02139, USA

† contribute equally

*mengkun.liu@stonybrook.edu

*db3056@columbia.edu

**Abstract:**

Superconducting transition-edge sensors are renowned for their extraordinary photon sensitivity and energy resolution, finding applications spanning quantum information, astronomy, and nanophotonics. Here, we report the development of BOlometric Superconducting Optical Nanoscopy (BOSON), a novel platform that integrates bolometric detection at the superconducting transition edges with near-field optical techniques. BOSON enables the mapping of photoinduced changes in superconductivity with unprecedented spatial resolution and photon sensitivity. By incorporating BOSON with low-dimensional materials, we achieved polariton imaging at nanowatt excitation levels—at least four orders of magnitude lower than the power typically required in prior near-field nanoscopy experiments. Our findings highlight the potential for BOSON to advance scanning probe based optical platforms to enable the detection of photons, polaritons, and Cooper pair dynamics at the nanoscale. This paves the way for quantum sensing applications using single-polariton detection and can offer deeper insights into quasiparticle dynamics.

**Introduction**

Superconductors (SC), celebrated for their unique electronic properties and extraordinary sensitivity to external perturbations, have long been integrated into cutting-edge quantum sensing technologies. Superconducting devices which respond to minimal stimuli enables innovations in single-photon detection (SPD) [1, 2, 3, 4, 5], bolometry [6, 7, 8, 9, 10, 11], and quantum sensing [12, 13]. For instance, superconducting bolometers leverage their small superfluid density [7] and pronounced change in electrical properties near the superconducting critical temperature ($T_c$) or current ($I_c$), rendering them highly responsive to minute photo-induced thermal perturbations. More recently, SPDs based on superconductor nanowires rely on the localized suppression of superconductivity upon absorption of a single photon, driving the nanowire over the critical current and contributing to a measurable signal [1]. Among various SC-based techniques, transition edge sensors (TESs) stand out for their extraordinary sensitivity and low noise. Operating at the brink of the phase transition, TESs function effectively across a broad range of photon energies independent of polarization [10, 13]. Recent advancements in TES-based superconducting detectors and SPDs, such as devices built on niobium nitride (NbN) [14], magnesium diboride ($MgB_2$) [3], and high $T_c$ superconductors [2] have expanded the operational temperatures, spectral coverage, and response speeds to new levels, enabling a wide variety of choices of SC-based devices for detecting weak light-matter interactions.

Concurrently, scanning probe optical nanoscopy, which uses tip-enhanced near-field light to probe or induce local electromagnetic responses, has emerged as a transformative tool for nanoscale optical imaging, spectroscopy, and nano-light maneuvering. For example, scanning near-field optical microscopy (s-SNOM) [15, 16], photocurrent nanoscopy [17, 18, 19, 20], and tip-enhanced nano-photo-lithography [21] have enabled many studies in complex material systems. In particular, photocurrent nanoscopy has emerged as a powerful technique for exploring nanoscale optical and photothermal effects [22]. For instance, graphene-based photocurrent nanoscopy, has been leveraged to probe local charge puddles [17] and plasmon polariton propagation in graphene [18], which utilizes graphene's exceptional Seebeck coefficient and low heat capacity [23] to detect local heat induced by polaritonic excitations. Due to the reduced electron-specific heat and electron-phonon interaction (phonon-cooling), the sensitivity of such systems can be improved at low temperatures. In such application, exceptional graphene quality and delicate device engineering are often required. Graphene bolometric effect is also explored to function as terahertz detector [24]. However, due to the weak temperature dependence of resistivity in bulk graphene, bolometric detection is challenging to be directly realized in DC transport measurements. These constraints significantly limit the applicability in detecting weak electromagnetic perturbations or resolving subtle light-matter interactions in a wide range of scenarios, where ultra-sensitive detection mechanisms are paramount.

Here we introduce BOlometric Superconducting Optical Nanoscopy (BOSON) as an ultra-sensitive scanning probe nanoscope that employs superconducting materials as on-chip bolometric detectors for photon and polariton sensing. We demonstrate how photocurrent near-field nanoscopy can be utilized to map local transition edge (TE) in superconducting thin films (e.g., Nb weak-link nanobridges), where local $T_c$ can be inferred. In addition, by integrating the Nb nanobridge junction with hexagonal boron nitride (hBN), we achieved hyperbolic phonon polariton imaging at nanowatt excitation levels—at least four orders of magnitude lower than the power typically required in a s-SNOM and graphene-based near-field photocurrent nanoscopy experiments. We believe these developments significantly enhance our capability to detect extremely weak nano-light signals with exceptional precision and a low-power threshold.

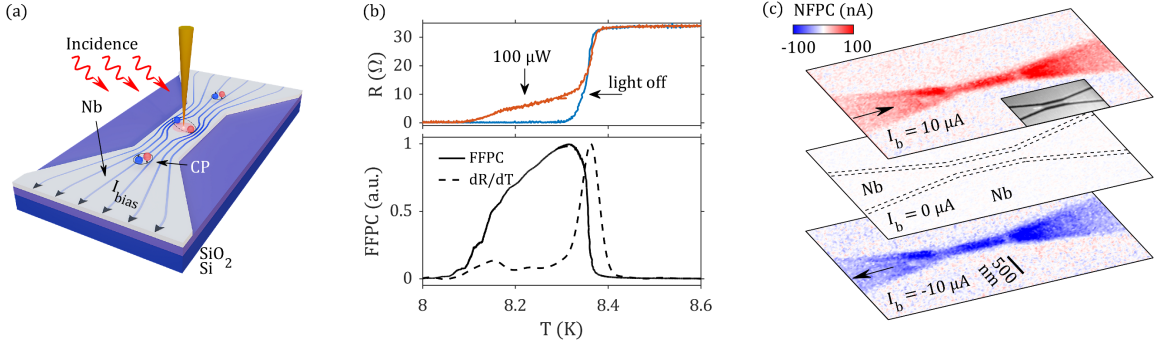

**Figure 1: Experimental setup and measurement concept of BOSON.** (a) Schematic of the experimental configuration, demonstrating the layered structure of the SC and the near-field photocurrent (PC) characterization. CP: Cooper Pair. (b) DC and far-field photocurrent (FFPC) characterization of an 8.5 μm wide, 50 nm thick Nb bridge device with an incident beam at 1555 cm$^{-1}$ and 100 μW. (c) Near-field bolometric photocurrent detection of a Nb nanobridge structure with a 200 nm-wide center bridge near $T_c$. Top: positively biased, middle: zero bias, bottom: negatively biased. The laser is at 880 cm$^{-1}$ and 50 μW. The double dashed line in the middle panel indicates the location for the lithographically defined trench that separate the active region and inactive region of the Nb film. The inset shows the AFM image of the Nb nanobridge structure.

**Fig. 1(a)** illustrates the typical configuration of a BOSON setup, integrating superconductors as photocurrent sensors for detecting tip-induced electromagnetic waves. The detection mechanism hinges on the suppression of super-carrier density by photo-excitation. The suppression drives a near-critical-point superconducting channel resistive, typically monitored at the steepest region of the resistance-temperature or resistance-bias curve, known as the transition edge [6, 7, 11]. The resulting resistive normal region generates a photocurrent or photovoltage signal under current bias, which is detected through the electrodes connected to the system. The photocurrent can be measured under two scenarios. One is the far-field photocurrent (FFPC) measurements, where we measure the photocurrent without the tip for basic sample characterization. The other is the near-field photocurrent (NFPC), where the tip-launched electromagnetic excitation induces a local change of sample properties. As a result, when the sample is raster-scanned under the tip, a tip-modulated photocurrent map can be generated. The NFPC enables the mapping of local sample properties under the tip apex, whereas the FFPC measures the photoinduced current across the entire sample ensemble. The details of both methods can be found in the supplemental materials. With this set up, we can measure transport properties, FFPC, and NFPC responses without major modifications to the system.

**Fig. 1(b)** shows the temperature dependence of the four-point resistance and FFPC of a 50 nm Nb film. The FFPC is measured with a ~10 μA bias current and a 100 μW incident illumination (1555 cm$^{-1}$, 192 meV). The signal is DC filtered, amplified and subsequently demodulated (measurement circuit is included in SI). All temperatures in this report are calibrated (calibration method is detailed in SI). Two distinct regimes are evident in the resistance measurements under laser illumination: a low-temperature regime where local heating occurs due to laser absorption, and a high-temperature regime corresponding to the global critical temperature ($T_c$). The FFPC measurements align with the derivative of the resistivity with respect to temperature, $dR/dT$, suggesting that the photocurrent response is mostly thermally induced within this experimental scheme.

Unlike graphene-based near-field photocurrent measurements, which rely on precise electrostatic gating near the charge neutral point [17], NFPC in BOSON experiments operates near the critical current of the working temperature. We utilize a specific type of samples for the NFPC measurements: a Nb thin film (50 nm thick) etched into a nanobridge junction where the current runs through a narrowed bridge at the center of the device. Here Nb is chosen for its robust and relatively high transition temperature as a conventional SC, and its simplicity of sputtering deposition and reactive ion etching patterning. The width of the bridge, 200 nm, is intentionally comparable to the coherence length of the Cooper pairs in Nb (50 ~ 100 nm) [25]. As a result, the bridge has a tiny local heat capacity, a limited superfluid density, and is extremely sensitive to external perturbations near the critical point.

The NFPC maps of the Nb nanobridge structure with a small bias applied in both directions are shown in the top and bottom panels of **Fig. 1(c)**. Without the bias current, no photocurrent signal is observed (middle panel in **Fig.1(c)**). The NFPC signal observed in our setup can, to the first order approximation, be considered proportional to the derivative of local voltage drop with respect to local temperature ($I_{loc} \cdot \Delta R_{loc}/\Delta T_{loc}$), where the local temperature change ($\Delta T_{loc}$) is induced by the tip field ($\Delta T \propto |E_{tip}^2|$). The tip-enhanced field is typically distributed in a range characterized by the tip apex size (20 nm) due to the lightning-rod effect [26, 20]. The distribution of local heating is subsequently determined by either quasiparticle diffusion ($L_D = \sqrt{D\tau_{th}}$) or thermal conduction ($L_T = \sqrt{\kappa\tau_{th}/C_e}$). Here, D is the diffusion parameter. $\tau_{th}$ is the thermalization time scale mainly limited by electron phonon coupling efficiency, $\kappa$ is the electron thermal conductivity and $C_e$ is the electron heat capacity. The typical thermal length scale is on the order of sub-microns to microns [7].

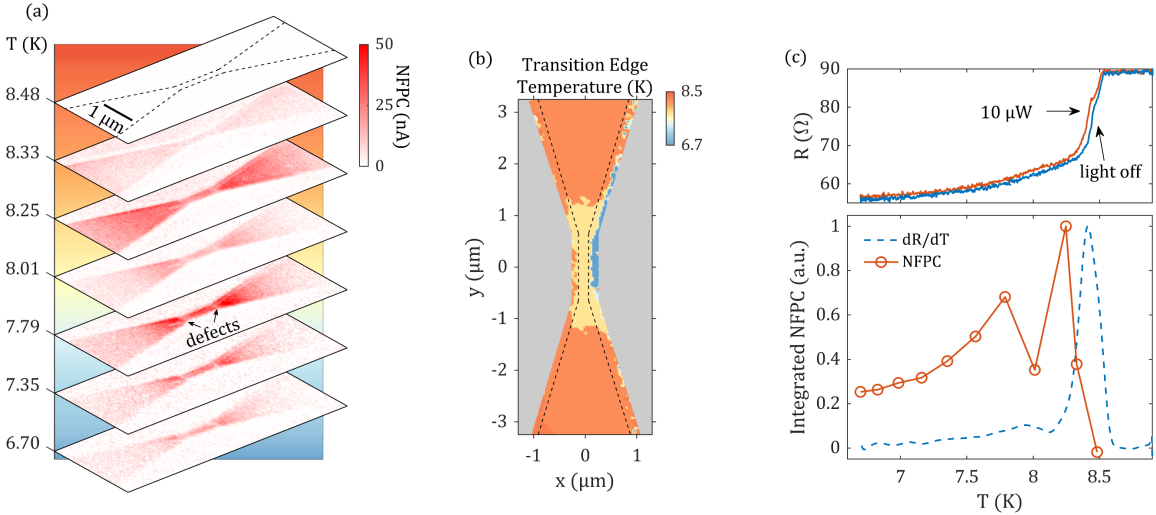

**Figure 2:** Temperature-dependent **transition edge mapping using BOSON.** (a) NFPC response of the Nb nanobridge structure same as in **Fig. 1(c)** at various temperatures. The structure is outlined by black dashed lines. The incident light is at 880 cm$^{-1}$ with 10 μW power, and the bias current is ~10 μA. (b) Transition edge temperature map of the Nb nanobridge, extracted from the data in (a). (c) The NFPC signal is integrated over each map in **Fig. 2(a)** and compared with the transport curve side-by-side.

Next, we demonstrate the application of BOSON for areal mapping of transition edges in superconductor devices. We report the NFPC measurements of the same Nb nanobridge structure in **Fig. 1(c)** with 10 µW incidence. At the lowest achievable temperature, the NFPC signal is predominately observed when the tip is located near or directly above the bridge, as seen in the 6.70 K panel in **Fig. 2(a)**. As the temperature increases, the signal at the bridge peaks at approximately 7.79 K and decreases beyond 8.01 K. Around 8.25 K, the NFPC signal extends across the entire SC sample, while the bridge region no longer yields the strongest signal. At 8.33 K, the NFPC signal over most of the scanning window begins to diminish and vanishes entirely at 8.48 K. Despite the tiny tip apex size (~20 nm), we notice that the generation of photocurrent signal is not confined to the size of the tip apex or the weak-link nanobridge but expands to the bulk of the sample. This is due to the finite diffusion or thermal length of quasiparticles at the micrometer scale. Upon tip-induced absorption, the micron-sized resistive region shuts down the superconducting channel, leading to an observable signal.

Local transition edge temperature under weak illumination and small bias current can be extracted from temperature-dependent BOSON images. In **Fig. 2(b)**, we map the spatially resolved transition edge temperature by locating the temperature where the NFPC signal is maximized. The mapping reveals a variation in TE temperatures between the bridge and other regions of the device, with the bridge exhibiting a lower temperature compared to the bulk of the sample. This spatially resolved behavior is consistent with the typical weak-link resistivity features measured in **Fig. 2(c)**. The bulk Nb region maintains an intrinsic TE temperature at $T_c$=8.5 K. The bridge region, however, has a lower TE temperature due to its comparable dimension scale to the coherent length of Nb superconductor. Notably, microscopic details of the junction become visible. For instance, two defects (labeled in **Fig. 2(a)**) appear in NFPC mapping at the top and bottom sections of the weak-link nanobridge, exhibiting lower NFPC responses [27]. The same measurement principle can be applied to examine the intrinsic and extrinsic inhomogeneity of other types of superconductors, such as two-dimensional (2D) SC $FeTe_{1-x}Se_x$ (FTS) thin flakes. A detailed discussion of the FTS results is provided in the supplemental materials.

Beyond probing local superconductivity, SC junctions can be integrated with low-dimensional van der Waals systems to enable ultra-sensitive BOSON mapping of diverse types of quasiparticles or elementary excitations. As a proof-of-principle demonstration, we showcase real-space mapping of polaritons with unprecedented sensitivity, which can be crucial for quantum information science as the confinement, propagation, and interactions of polaritons have enabled many interesting observations [15, 16, 28, 29, 30, 31] or proposals of applications in correlated quantum systems [32, 33].

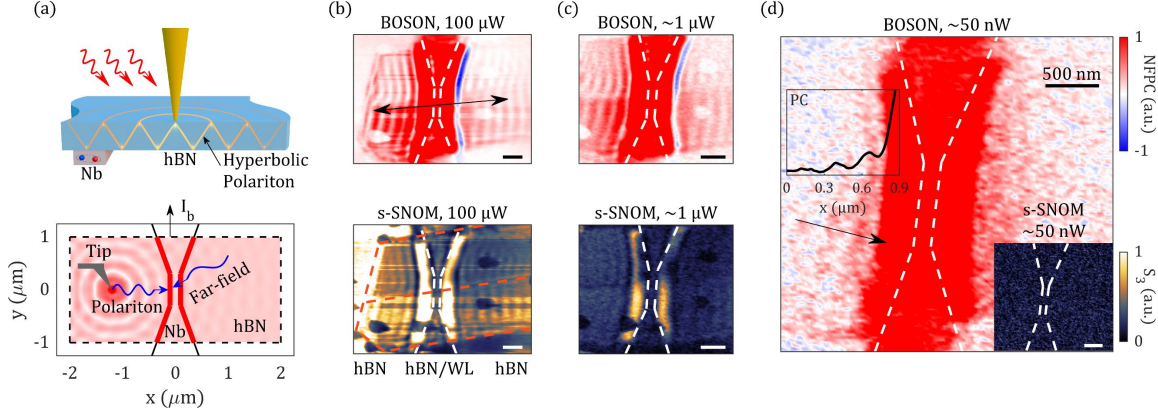

**Figure 3: On-chip nanowatt sensing of polaritons with current-induced transition edges.** (a) Top panel: Schematic representation of polariton sensing using an on-chip TES. Bottom panel: A potential mechanism of polariton sensing. The tip-launched polaritons interfere with the strongly enhanced local field near the Nb edges. These Nb edges become a strong thermal absorber, making the weak link a sensitive detector to polariton wave fronts. The detailed mechanism and simulation are summarized in SI. (b, c) Identification of hyperbolic phonon polaritons through BOSON imaging (top) and s-SNOM imaging (bottom) with 100 μW and ~1 μW incident power, respectively. BOSON and s-SNOM images are recorded simultaneously. The white dashed lines demarcate the Nb weak-link nanobridge structure, while the red dashed lines outline the hBN flake with a thinner upper region and a thicker lower region. The black arrow marks the location where the bias-sweep signal in the following discussion is collected. WL: weak-link. Scale bar: 500 nm. (d) BOSON imaging of hyperbolic phonon polaritons with incident light power as low as ~50 nW. The black arrow indicates a line-cut, revealing the fringe pattern in the top left inset. The bottom right inset shows that s-SNOM measurements fail to yield meaningful results under the same ~50 nW illumination.

We leverage the Nb-based nanobridge structures to enhance the detection sensitivity of polaritons in hBN, a 2D van der Waals material renowned for its hyperbolic phonon polaritons [28]. The hyperbolic phonon polariton tracks a unique propagation path in the hBN flake, as shown in **Fig. 3(a)**. The heterostructure was prepared through exfoliation and stacking of a ~50 nm thick hBN flake onto a similar device used in **Fig. 2(a)**, with a Nb bridge of 250 nm width in the center (see supplemental materials). In this experiment, a QCL laser tuned to mid-infrared frequencies at 1491.9 cm$^{-1}$ is used to excite polaritons in the hBN layer. Two types of near-field images are simultaneously acquired: the NFPC signal from BOSON measurements and the tip-scattered signal ($S_3$) from conventional s-SNOM for comparison. The basic principles [34] and specific parameters used in our s-SNOM measurements are detailed in the supplemental materials.

Upon optical excitation, polariton waves are excited by both the tip and sharp edges of the structure. In conventional s-SNOM imaging of hBN polaritons [28], two types of fringe pattern can be observed at different locations of an hBN flake. Neighboring metallic boundaries, edge-launched polaritons contribute to a dominating fringe pattern with an intrinsic polariton wavelength $\lambda = \lambda_p$. Near the natural boundary, tip-launched polaritons contribute to both $\lambda = \lambda_p$ and $\lambda = \lambda_p/2$ patterns depending on whether the scattering happens at the boundary or the tip [28]. Both types of polariton pattern can be observed in the s-SNOM channel as analyzed in the supplemental materials. In contrast, the PC signal detected using BOSON originates solely from tip-launched polaritons reaching the Nb nanobridge (bottom panel of **Fig. 3(a)**). The observed fringe pattern results from the interference between tip-launched polaritons and strongly enhanced local field near the Nb edges. We explain the detail of the potential mechanism of polariton sensing using BOSON method in the supplemental materials. We confirm that the experimentally extracted and calibrated wavelength of the fringe pattern in the PC channel is $\lambda = 268$ nm in the thicker hBN (bottom half flake) and $\lambda = 231$ nm in the thinner hBN (top half flake). The wavelengths linearly depend on the thickness of flakes [28]. The two wavelengths correspond to polariton modes screened by a metal (Nb) substrate with hBN thicknesses of 59 nm and 51 nm, consistent with our calibrated AFM measurement. As the tip sweeps across the sample, the nanobridge experiences a variation of localized electromagnetic field due to field interference, leading to changes in NFPC signal generation. In **Fig. 3(b)** to **Fig. 3(d)** we record the NFPC and s-SNOM signal under different incident laser power (100 µW, 1 µW and 50 nW). Experimental data confirm robust photocurrent imaging of hyperbolic phonon polariton fringes even under optical excitation as low as ~50 nW, whereas in s-SNOM measurements, no trace of the signal can be seen with this low incident energy (inset of **Fig. 3(d)**). Therefore, these results demonstrate BOSON's unprecedented sensitivity in probing local elementary excitations or quasiparticles induced by nanoprobes. As a comparison, previous near-field optical methods typically require milliwatt to sub-milliwatt power levels to generate detectable signals [15, 16, 28, 35, 17, 19, 20, 36, 37, 38].

We note that the resolution of BOSON measurements for sensing polariton waves is strongly influenced by the geometric design of the bridge structure. The constriction leads to a concentration of supercurrent density within the bridge region. By finely tuning the temperature and bias current, the signal originating from the bridge can be made to dominate the overall response. As a result, the spatial resolution perpendicular to the bridge is effectively set by the bridge width (see supplemental materials). In the current design, a 250 nm-wide bridge is sufficient to resolve the hyperbolic polaritons in hBN. In optimized structures, both the width and depth of the bridge are expected to be reduced to much smaller dimensions—down to 50 nm—thereby enhancing BOSON's capability to resolve finer polaritonic features.

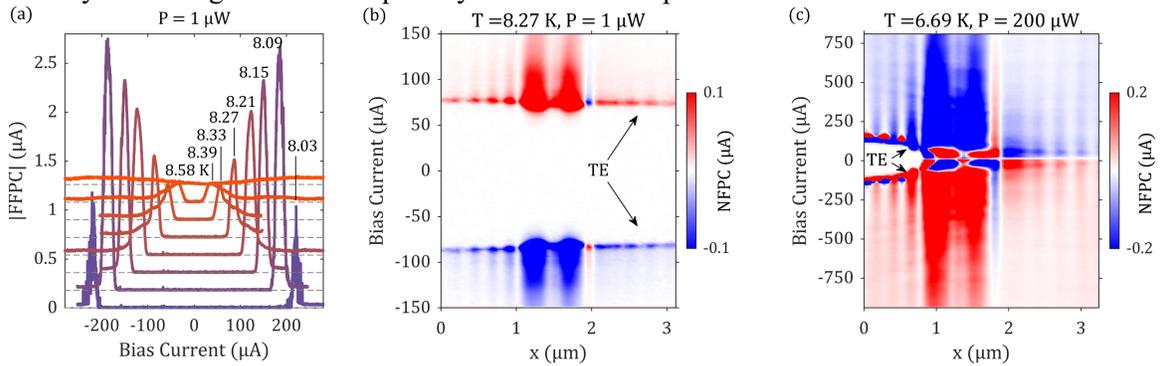

**Figure 4: On-chip detection of far-field radiation and near-field polariton.** (a) FFPC of the Nb+hBN junction under 1491.9 cm$^{-1}$ and 1 µW incident power. (b) NFPC linescans at T = 8.27 K under varying bias current with weak illumination P = 1 µW. TE: transition edge. (c) NFPC linescans at T = 6.69 K with strong illumination P = 200 µW.

The optimal working condition for maximized responsivity can be determined by analyzing the temperature- and current-dependent photocurrent map. This process is illustrated by a series of FFPC measurements with 1 µW illumination at various working temperatures (**Fig. 4(a)**). At 8.58 K, where the weak-link nanobridge is slightly above the superconducting transition $T_c$, only a mild photocurrent is observed when sweeping the bias current. Below $T_c$ (<8.5 K), no photocurrent is detected at bias currents lower than the critical current $I_c(T)$. However, beyond this critical current, a pronounced and narrow peak is observed, followed by a mild response at larger bias currents. As the operating temperature decreases, the critical current $I_c(T)$ increases, and the peak values roughly follow a linear trend $PC_{max}=I_c(T)R_N$ [7, 11], where $R_N$ is the resistance of the Nb nanobridge in the normal state. At 8.09 K, the photocurrent reaches its highest measurable value achievable in our setup. Below 8.09 K, we observed excess noise in the FFPC signal. The large FFPC at superconductor-normal switching can be understood by estimating the bolometric response to be proportional to $\frac{dV}{dI}\Big|_{I_c} \frac{dI_c}{dT}$. A steep slope in the current-voltage relation leads be a giant bolometric response. At the low temperature end of the switching, however, the resistance switching under a large current bias becomes abrupt and non-continuous due to rapid thermal-runaway effect. As a result, $\frac{dR}{dI}\Big|_{I=I_c}$ diverges and experimentally a large noise emerges. This issue may be resolved with a voltage-biased (hence with negative feedback) measurement and improvements in low-noise electronics. Combining with a reduction of the Nb thickness, optimization of weak-link bridge geometry, and enhancements in material properties, we anticipate that photocurrent sensitivity could be improved by orders of magnitude, potentially even enabling the detection of polaritons at the single photon level [2, 3].

With the knowledge of the $I_c(T)$ under light illumination, we characterize PC sensitivity near the critical bias for detecting the tip-launched polaritons. In **Fig. 4(b)** and **4(c)**, we present an NFPC line scan across the junction under varying bias currents, taken along the black arrow indicated in **Fig. 3(b)**. With weak illumination power (**Fig. 4(b)**), the polariton features, together with other NFPC responses, are absent with a current $< I_c(T)$. Polaritonic fringes emerge with the highest sensitivity precisely at the transition edges for both positive and negative bias currents. As the bias current increases beyond the transition edges, the polaritonic fringes gradually diminish in the NFPC signal.

Under high illumination power and lower operating temperatures (**Fig. 4(c)**), a distinct behavior emerges. The critical bias required to observe polaritons becomes location-dependent and exhibits asymmetry depending on whether the tip is positioned on the left or right side of the bridge. This asymmetry arises from a slight difference in the far-field light illumination received by the weak-link nanobridge. At high illumination power, this effect becomes pronounced: on the right side, the combination of far-field illumination and the tip-launched local field drives the weak-link nanobridge above the superconducting transition even at minimal bias current. In contrast, the left side requires a larger bias current to reach the transition edge. Additionally, we observed a more complex sign-switching behavior of the photocurrent signal compared to the expected polarity switching following the bias current. This phenomenon is likely attributed to the strong nonlinearity of the resistance-temperature relationship $R(T)$ near the TE, leading to additional sign-switching effects.

**Summary:**

By integrating superconducting weak-link nanobridge structures into s-SNOM, we have developed BOSON, a superconducting-based nanoscopy platform capable of resolving superconducting transitions at the nanoscale and probing collective quasiparticle excitations with exceptional sensitivity. This approach leverages the bolometric response in superconductors—a highly sensitive detection mechanism based on perturbations to the superconducting condensate. Compared to existing s-SNOM and photocurrent nanoscopy techniques, this method achieves significantly enhanced sensitivity and operates at nanowatt power levels, effectively minimizing sample and tip heating. This capability facilitates the exploration of quantum phenomena at extremely low temperatures with reduced noise, establishing BOSON as a superior tool for ultrasensitive nanoimaging applications, particularly for weak bosonic excitations such as Cooper pairs and polaritons. It also enables the study of low-energy excitations in the THz frequency range, offering new nano-optical functionalities based on existing bolometric techniques.

These advancements enable optical nanoscopy measurements using potentially a wide range of low-fluence light sources such as globar and synchrotron-based facilities [38, 39]. The capability for reporting local TE temperature can be critical for characterizing superconductor devices such as Nb- or Ta-based capacitors in solid-state qubits, where surface defects, oxidation, and metal-proximity effects are known to impact coherence time [40, 41, 42]. Moreover, BOSON holds the potential to address long-standing challenges in quantum nanophotonics, such as single-polariton detection. Recent breakthroughs in high critical temperature materials such as BiSrCaCuO [2], LaSrCuO [2] and $MgB_2$ [3] in quantum sensing raise questions about the microscopic photon detection mechanism. The complex electronic band structures and larger superconducting energy gaps in high-$T_c$ materials challenge traditional methods routinely used for modeling the detection in low-$T_c$-based sensors. Furthermore, it also remains unclear how irradiation-induced defect formation, a crucial method for high-$T_c$-based sensors, contributes to the improved performance observed after irradiation.

Building on the BOSON framework, future work will aim to extend the capabilities of superconductor-based nanoscopy, covering a broader frequency range, unveiling local superconductivity dynamics, and potentially achieving Single Photon/Polariton Imaging Nanoscopy (SPIN).

**Author contributions:**

R.J., B.Z. and M.K.L. conceived the project and designed the experiments. G.L.C., I.C., D.B., Q.L., K.K.B., D.N.B., X.D. and M.K.L. supervised the project. B.Z., D.K. and X.D. prepared the devices. R.J. and B.Z. performed the experiments with the support of W.Z., Z.Z., H.W., B.C., J.H.P., L.W., Z.D., S.C and C.D.P. X.C. performed the finite-element simulation. R.J., B.Z. and M.K.L. analyzed experimental data with the input from X.D., G.L.C., I.C., D.B., K.K.B. and D.N.B. R.J., B.Z. and M.K.L. cowrote the manuscript with input from all co-authors.

**Acknowledgements:**

The research of effective near-field quantum sensors is primarily supported by Programmable Quantum Materials, an Energy Frontier Research Center funded by the U.S. Department of Energy (DOE), Office of Science, Basic Energy Sciences (BES), under award DE-SC0019443. R. J., Q. L., and M. K. L. acknowledge support from the U.S. Department of Energy, Office of Basic Energy Sciences, Division of Material Science and Engineering, under contract No. DE-SC0012704 for the support of studies of superconductor-based quantum transport system. D.N.B, M.K.L., L.W. acknowledge funding of novel material characterization by the U.S. Department of Energy, Office of Science, National Quantum Information Science Research Centers, Co-design Center for

Quantum Advantage (C2QA) under contract number DE-SC0012704. M.K.L. acknowledges the Gordon and Betty Moore Foundation DOI: 10.37807/gbmf12258 for supporting the development of polaritonic materials. M.K.L. acknowledges support for developing sensitive probes for electrodynamics in chiral materials from the NSF Faculty Early Career Development Program under Grant No. DMR – 2045425. X.D. acknowledges support from the US Army Office of Research (ARO) under award: W911NF2410148. D.A.B. acknowledges the support of Singapore Ministry of Education Tier 2 grant award T2EP50123-0020. We also acknowledge Prof. Chunjing Jia for helpful discussions.

**Data and materials availability:** The data that support the findings of this study are available from the corresponding author upon reasonable request.

# Bolometric Superconducting Optical Nanoscopy (BOSON)

Ran Jing[1,2,†], Boyi Zhou[1,3,†], Dingchen Kang[1], Wenjun Zheng[1], Zijian Zhou[1], Heng Wang[1], Xinzhong Chen[1], Juntao Yao[2,4], Bing Cheng[1], Ji-Hoon Park[1], Lukas Wehmeier[6], Zhenbing Dai[1], Shoujing Chen[1], Christopher D. Prainito[1], G. L. Carr[6], Ilya Charaev[7], Denis Bandurin[8], Genda Gu[2], Qiang Li[1,2], Karl K. Berggren[9], D. N. Basov[5*], Xu Du[1], Mengkun Liu[1,6*]

[1] *Department of Physics and Astronomy, Stony Brook University; Stony Brook, New York 11794, USA*

[2] *Condensed Matter Physics and Materials Science Division, Brookhaven National Laboratory; Upton, NY, 11973-5000, USA.*

[3] *Center for Integrated Science and Engineering, Columbia University, New York, New York 10027, USA*

[4] *Department of Materials Science and Chemical Engineering, Stony Brook University, Stony Brook, New York, 11794 USA*

[5] *Department of Physics, Columbia University; New York, New York 10027, USA.*

[6] *National Synchrotron Light Source II, Brookhaven National Laboratory; Upton, New York 11973, USA.*

[7] *Physics Institute, University of Zurich, Zurich 8057, Switzerland*

[8] *Department of Materials Science and Engineering, National University of Singapore, 117551, Singapore*

[9] *Department of Electrical Engineering, Massachusetts Institute of Technology, Cambridge, MA 02139, USA*

†contribute equally

*mengkun.liu@stonybrook.edu

*db3056@columbia.edu

**Contents**

1. **Methods for far-field and near-field measurements.**

The near-field experiment of Bolometric Superconducting Optical Nanoscopy (BOSON) and scattering-type near-field scanning probe microscopy (s-SNOM) is performed on our home-built near-field microscope which can work down to T ≈ 5 K (sample temperature) and up to B = 7 T [1, 2]. In the experiment, an incident beam is focused on an Au-coated Akiyama AFM tip (NanoAndMore USA Corporation). The electric field is strongly enhanced on the tip apex (radius ~20 nm) due to the lightning-rod effect. The AFM works in tapping mode with the tip oscillating at a frequency Ω ≈ 50 kHz. For s-SNOM measurements, the metallic tip scatters the near-field electrodynamics to far-field radiation, which is subsequently collimated and focused into an MCT photodetector. Both the near-field scattering signal (s-SNOM signal) and the near-field photocurrent signal (BOSON signal) are encoded in integer harmonics of the tip oscillation frequency. To suppress unwanted far-field artifacts, we extract the 3rd harmonic of the near-field scattering signal ($S_3$ at 3Ω) and the 1st or 2nd order photocurrent signal (NFPC at Ω or 2Ω).

The far-field photocurrent (FFPC) measurement is acquired with the tip retracted and the beam focusing on the sample surface. The beam is modulated at a frequency of 300~1000 Hz using an external chopper. A lock-in amplifier, synchronized to the chopping frequency, extracts the change of DC current by demodulating the response at the fundamental frequency of light modulation.

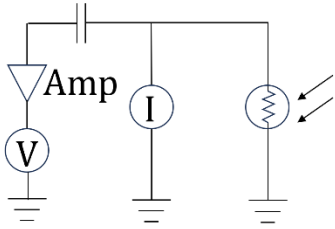

**Fig. S1: Photocurrent measurement circuit.**

The photocurrent signal is measured using the circuit shown in **Fig. S1**. A Keithley 2400 is used as a current source with a 2 kΩ resistor. In future experiments, the current source can be replaced with a low-noise voltage source for better signal to noise performance. The DC signal is frequency-filtered before being amplified. The amplified signal is demodulated with a Zurich lock-in amplifier.

2. **Type-II hyperbolic phonon polariton of hBN**

Phonon polaritons reside in the Reststrahlen band of a strong phonon resonance. In isotropic or weakly anisotropic materials, the isofrequency surface of phonon polariton is spherical or ellipsoidal. Beams of polaritons or waveguide modes travel in all directions. In strongly anisotropic materials like hBN, in-plane and out-of-plane phonons reside in drastically different frequencies. The primary in-plane phonon spans from 1360 cm$^{-1}$ to 1614 cm$^{-1}$, while the primary out-of-plane phonon spans from 760 cm$^{-1}$ to 825 cm$^{-1}$. In the range of in-plane phonon frequencies, the in-plane dielectric constant is negative, while the out-of-plane dielectric constant is positive [3]. The opposite signs in dielectric tensor contribute to hyperboloidal isofrequency surface of polariton mode [4]. The phase and group velocity of the polariton is limited to a range of angles with respect to the crystal axis. At the high momentum limit, polariton modes asymptotically approach fixed velocities, contributing to high mode density. As a result, polaritons traveling in this frequency range mainly propagate at a specific angle calculated by $\tan\theta = \sqrt{-\epsilon_\parallel/\epsilon_\perp}$, where $\epsilon_{\perp/\parallel}$ are in-plane (perpendicular to crystal axis) and out-of-plane (parallel to crystal axis). A typical beam traveling pattern is shown in **Fig. 3(a)**. The wavelength scales linearly with the thickness of the flake.

The wavelength of hBN polariton can be computed at the frequency (1492 cm$^{-1}$) used in the measurement of Fig. 3. If the 50 nm hBN flake is placed on top of a dielectric substrate, the computed wavelength is around 696 nm. In our experiment, the flake is placed on top of Nb. The Nb serves as a metallic boundary, contributing to additional screening of the mode. The wavelength of the hBN polariton is then computed to be 230 nm with incident light frequency at 1492 cm$^{-1}$.

### 3. Fabrication of the Nb nanobridges and Nb-hBN junctions.

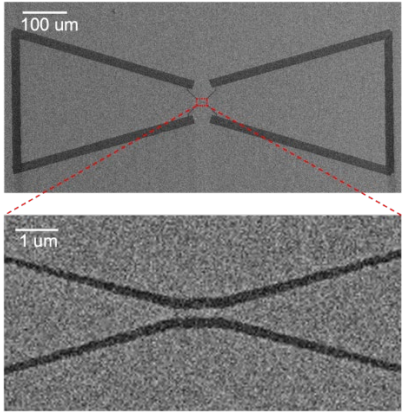

**Fig. S2: SEM image of Nb nanobridge.**

The fabrication of Nb bridges begins with the deposition of a high-quality Nb thin film onto a Si/SiO$_2$ substrate using DC magnetron sputtering in argon atmosphere. In our deposition system, the argon pressure is set to 8.8 mTorr under a sputtering power of 400 W, to achieve optimized superconducting transition temperature and minimized stress. The sub-micron bridge structures are then defined using electron-beam lithography (EBL) followed by reactive ion etching (RIE) to create narrow constrictions connected to wiring pads, both isolated from the rest of the Nb film (See Fig. S2). The room-temperature RIE process for Nb uses a 25W SF$_6$ plasma at a pressure of 15 mTorr, which yields an etching rate of ~1nm/sec. The RIE definition of the devices, as opposed to the commonly used lift-off process, avoids sidewalls at the edges of the patterns and ensures a smooth surface critical for integration of hBN crystals and the scanning probe measurements. To ensure a clean surface prior to the subsequent hBN transfer, the Nb junction is subjected to a solvent cleaning process to remove residues and adsorbates that could interfere with interfacial quality.

We transfer the thin flake of hBN on top of the Nb nanobridge to form the superconductor detection junction following the standard dry transfer method [5]. The hBN flake is exfoliated onto a Si/SiO$_2$ substrate and then transferred onto Nb nanobridge using a polycarbonate (PC) film over a polydimethylsiloxane (PDMS) stamp. The stamp expansion and retraction speeds are carefully controlled to ensure minimal wrinkling and strain. Atomic force microscopy (AFM) is employed to verify the uniformity of the hBN coverage and ensure that surface roughness remains minimal, thereby preserving the intrinsic superconducting properties of the Nb nanobridge and facilitating optimal electronic coupling at the Nb/hBN interface.

### 4. Temperature calibration

In this experiment, all temperatures reported are calibrated based on the readings in our home-built magnetic cryo-SNOM. The raw reading $T_{raw}$ deviates from the actual sample temperature $T_{sample}$ by a few

Kelvins below 10 K. The calibration relies on the predominant superconductivity transition of the bulk of samples (regions away from the weak-link nanobridge). The transport properties of investigated samples are measured in a separate dilution fridge which cools down to 330 mK. The dilution fridge has better thermal shielding and no external optical illumination is introduced. Therefore, the reading of the predominant transition, $T_{SC}$ = 8.5 K for Nb with a thickness of 50 nm acquired in the dilution fridge is the first anchor point of the calibration. For Nb with thinner thickness, the transition temperature should be lower than 8.5 K. The 6.7 K transition temperature of a 20 nm thick sample is the second anchor point of the calibration.

We assume the deviation between $T_{raw}$ and $T_{sample}$ follow three fundamental features. First, in the high-temperature limit ($T_{raw} \geq 20$ K), $T_{sample}$ should equal $T_{raw}$. Second, approaching the base raw temperature, the sample temperature asymptotically approaches an elevated temperature $\Delta T$ above the base raw temperature. Third, in the transport curve, both anchor points should be satisfied. We use a simple relation to characterize the temperature deviation,

$$T_{sample} = (T_{raw}^{\alpha} + \Delta T^{\alpha})^{1/\alpha}$$

With two anchor points, we have $\alpha$=1.398 and $\Delta T$~5.224 K.

## 5. BOSON measurements of FeTe$_{1-x}$Se$_x$ (FTS) film

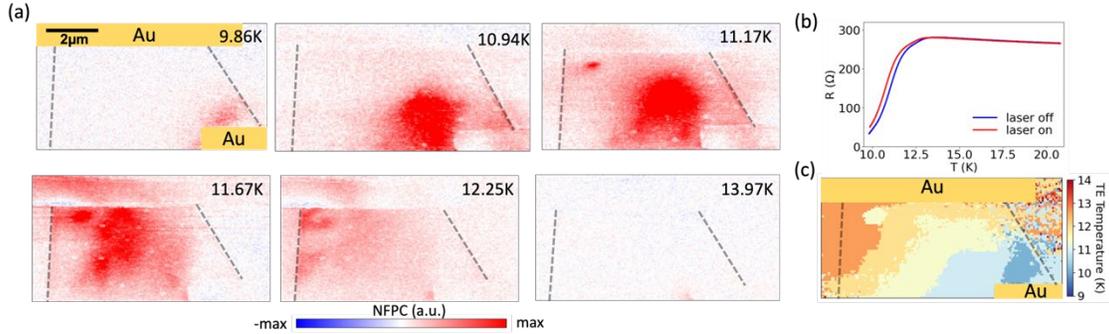

**Fig. S3: BOSON mapping of FeTe$_{1-x}$Se$_x$ micro flakes.** (a) The NFPC response and map of the transition edge temperature of an FTS micro flake. (b) Temperature-dependent resistance of the FTS with and without incident light. (c) Transition edge temperature map of FTS, as extracted from the experiments in (a).

FTS is an iron-based superconductor that has garnered significant attention recently due to its intriguing electronic properties [6, 7, 8, 9, 10], unconventional superconductivity [11, 12, 13, 14], and potential applications in quantum technologies [15]. As part of the iron chalcogenide family, FTS exhibits a tunable $T_c$ that varies with selenium content, reaching ~14.5 K for optimal compositions. This tunability is accompanied by a suppression of antiferromagnetic order and the emergence of robust superconducting phases. As a result, FTS is expected to display spatial variations in superconducting properties, including local $T_c$ and quasiparticle dynamics due to changes in selenium content [16, 17]. These inhomogeneities may be of significant interest, as they provide insights into the interplay between electronic structure and superconductivity at the nanoscale [18, 19, 20, 21]. Here we showcase the BOSON measurements of FTS, demonstrating the spatial variation of the local transition edge (TE) temperature can be readily obtained.

The FTS sample is cleaved into a 10 $\mu$m by 5 $\mu$m, 20 nm thick micro flake, which yields a $T_c$ of about 12 K (in the absence of light illumination). Current is measured through the two electrodes positioned on the sample, as illustrated in the top-left panel of Fig. S3(a). At a temperature significantly lower than $T_c$ (9.86 K), the NFPC signal appears only at the bottom-right region of the field of view, indicating that the local TE temperature is lower. As the temperature increases, the transition edge region shifts progressively from the lower electrode toward the higher electrode and disappears entirely above 13.97K. This series of temperature-dependent mapping in Fig. S3 (a) highlights the microscopic variation of local SC properties in an area defined between the electrodes. It also suggests that the broadened temperature range of the superconductor transition in this device (Fig. S3(b)) comes from the local $T_c$ inhomogeneities, which can now be clearly identified. Here, we report at a length scale of the phase inhomogeneities that is much smaller than previous studies using energy-dispersive X-ray spectroscopy [16]. The mapping of the transition edge temperature can be extracted from the NFPC mapping and is plotted in Fig. S3(c).

### 6. Magnetic field-dependent BOSON imaging of Nb nanobridge and FTS.

The magnetic-dependent mappings of the local TE in Nb nanobridge (the same sample as in Fig. 2) at 8.06 K are shown in Fig. S4. The magnetic field effectively reduces $T_c$ therefore decreasing the TE temperature in Nb sample. As field increases, the TE temperature approaches the sample temperature, leading to an enhancement of the signal. The signal reaches the maximum at around 0.20 T and decreases with further increase in the field, which moves the TE temperature away from the sample temperature. When the field exceeds the critical field, the sample transitions to the normal state, causing the signal to vanish completely.

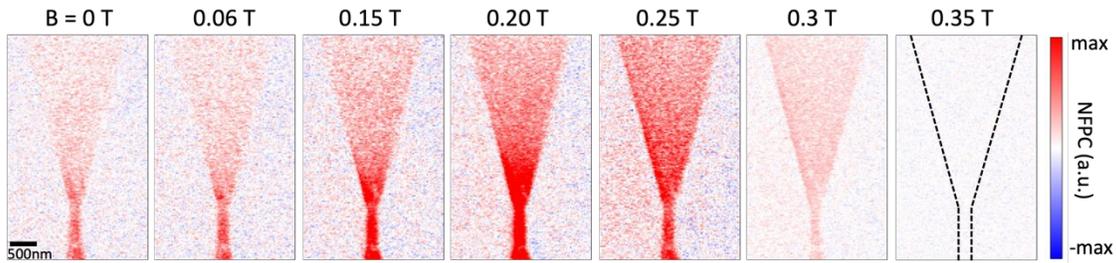

**Fig. S4: Magnetic field-dependent BOSON mapping of the Nb nanobridge at 8.06 K.** The magnetic field gradually suppresses $T_c$, leading to the shifts of the superconducting transition edge.

Fig. S5 shows the magnetic-dependent mappings of the local TE in the FTS sample. By increasing magnetic field, the transition becomes broader and $T_c$ shifts towards lower temperatures. At a temperature of ~9.86 K, which is below $T_c$ at 0 T, the applied magnetic field brings the TE temperature closer to the sample temperature and enhances the overall signal. However, at ~11.4 K which is near the $T_c$ at 0 T, the signal decreases with increasing magnetic field as the TE temperature shifts away from the sample temperature. In contrast to Nb, the critical field of FTS film is larger than 30 T [22]. Therefore, a magnetic field of 7 T will not cause the disappearance of signal in FTS.

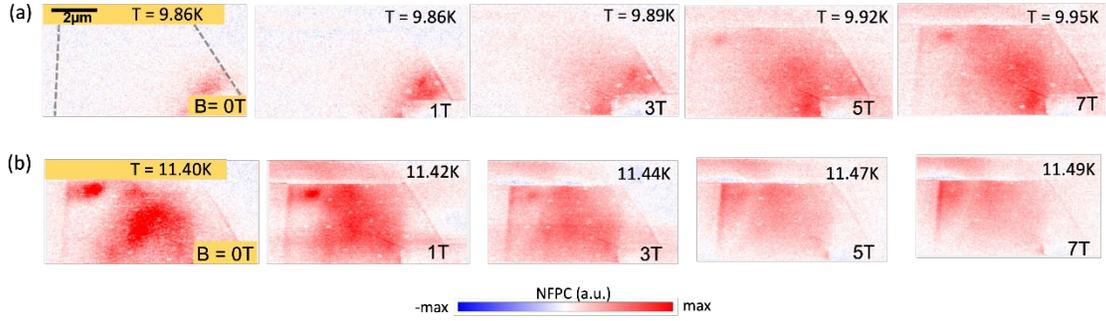

**Fig. S5 Magnetic field-dependent BOSON mapping of the FTS at two different temperatures.** (a) At lower temperatures (~9.86 K), the magnetic field leads to the overall enhancement of the NFPC in the field of view as the transition edge moves closer to the measurement condition. (b) At higher temperatures (~11.4 K), the magnetic field leads to the overall decrease of the NFPC in the field of view as the transition edge moves further away from the measurement condition. We note that a small temperature up-shift accompanies the application of the magnetic field.

## 7. Hyperbolic Polariton observed in s-SNOM image

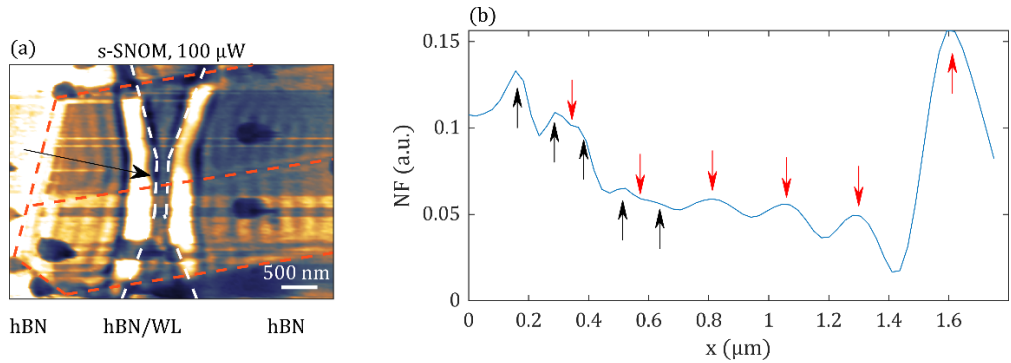

**Fig. S7: Polariton patterns in s-SNOM image.** (a) s-SNOM image of hBN hyperbolic polariton. (b) A linecut, marked by black arrow, is extracted from panel (a). The result is shown in panel (b). Black arrows indicate $\lambda_p/2$ fringes and red arrows mark the $\lambda_p$ fringe.

In s-SNOM measurements, two types of polariton patterns can be observed. Near metal boundaries, edge launched polaritons are scattered by the tip, resulting in $\lambda = \lambda_p$ fringes. Near natural boundaries, tip launched polaritons are scattered by the boundary, resulting in $\lambda = \lambda_p$ fringes as well. In the meantime, the tip launched polaritons are also reflected by the boundary and subsequently scattered by the tip. The doubled travel path produces $\lambda = \lambda_p/2$ fringes.

In our experiment, we can observe both types of polariton pattern in s-SNOM images. As an example, we make a linecut in the image with 100 µW illumination (Fig. S7(a)). The linecut goes across a natural hBN boundary and a Nb edge. In Fig. S7(b), $\lambda = \lambda_p/2$ and $\lambda = \lambda_p$ fringes are marked with black and red arrows, respectively. We notice near the natural boundary on the left, $\lambda = \lambda_p/2$ fringes overlay on top of $\lambda = \lambda_p$ fringes. Near the Nb edge on the right, a strong $\lambda = \lambda_p$ pattern is observed.

## 8. Simulation of polariton detection using BOSON method

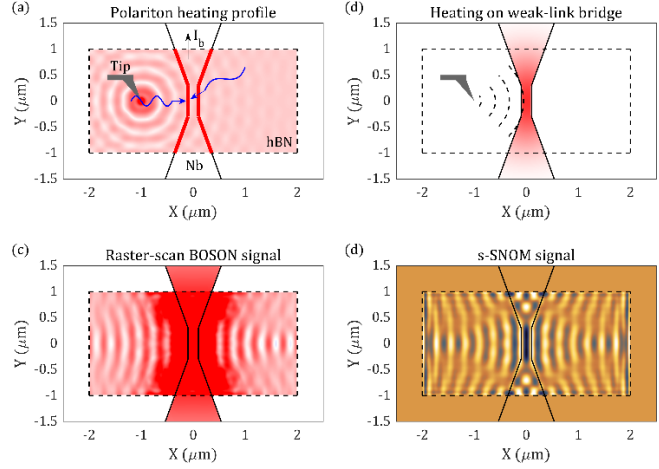

**Fig. S8: Simulation of polariton detection using BOSON method.** (a) With optical illumination, tip initiates a polariton propagation on the sample surface. A wave-like heating profile is generated due to the interference between the polariton field and far-field illumination. (b) Tip-initiate polariton waves reach Nb structure, causing heating effects on the weak-link bridge. Due to quasiparticle diffusion and thermal conduction, the plasmon wave-like heat profile is smeared out. Here the characteristic diffusion length is assumed to be 0.5 μm. (c) Sweeping the tip across the sample surface, we calculate the PC signal based on panel (b) for each tip location. Consequently, we can achieve the BOSON polariton profile. (d) The s-SNOM signal is simulated based on three contributing channels: tip-launch-edge (hBN) reflect-tip-scatter, edge (Nb) launch-tip-scatter and tip-launch-edge (Nb) scatter.

We simulate the mechanism of polariton detection in BOSON using a simplified model that captures the essential photothermal response. Upon optical illumination, the metal-coated tip launches a circular polariton wave centered at the tip location. The polariton field at position $r$ is given by,

$$\tilde{E}_p = \frac{1}{a_{tip} + |r - r_{tip}|} E_0 e^{-\frac{|r - r_{tip}|}{l_0}} e^{ik_p \cdot (r - r_{tip})}$$

Where $k_p$ is the polariton wavevector. $l_0$=3 μm is the damping length. The $\frac{1}{a_{tip}+|r-r_{tip}|}$ factor captures geometric spreading of the wave. Here the radius of the tip is $a_{tip}$=20 nm. The polariton field causes a periodic heating profile [Fig. S8(a)] when interfering with the far-field illumination $\tilde{E}_F = E_{F0} e^{i(k'_f \cdot r + \phi_0)}$ and field caused by the Nb edges $\tilde{E}_{Nb}$,

$$P \propto \left| \tilde{E}_p + \tilde{E}_F + \tilde{E}_{Nb} \right|^2$$

The $k'_f$ represents the projected in-plane component of the free-space wavevector, and the $\phi_0$ represents the phase offset between the far-field illumination and the polariton field. In the experiment and the simulation, the far-field illumination track along the -x direction with an 30° out-of-plane pitch angle. The phase offset is assumed to be 0. Both parameters do not significantly impact the simulation result.

We note that Nb edges produce a strong enhancement of local field, and the resulting edge launched polariton upon optical illumination. These fields are accounted by the term $\tilde{E}_{Nb}$. The tip launched polariton would interfere with both fields and cause heating effect on the hBN flake. If the heating effect takes place on top of the Nb weak-link structure, a finite signal can be induced. As a result, these edges act as localized absorbers. In another word, these edges play a dominant role in detecting the polariton wave front. Upon thermal absorption, quasi-particle diffusion and thermal conduction thermalizes a broader area of weak-link bridge. In this simulation, we assume the characteristic thermalization length scale $l_D$=0.5 μm. The resulting thermal profile of the Nb structure is shown in Fig. S8(b). Due to thermalization or diffusion, regions away from the narrow bridge can also function as an efficient detector to polariton wave fronts.

To extract the BOSON signal, we compute the resulting photocurrent from the spatially varying temperature distribution:

$$\delta I = \int j_{bias}(r) \cdot \delta R(T(r)) \, d^2 \boldsymbol{r}$$

where $j_{bias}$ is the local bias current density, and $\delta R(T)$ is the resistance change due to local heating. Raster-scanning the tip yields a spatial BOSON response [Fig. S8(c)] that reflects the polariton interference fringes. We note that, when the tip parks on regions outside the hBN flake, a temperature profile characterized by the thermalization length scale is also generated, causing a significant photocurrent signal.

In Fig. S8(d), we simulate s-SNOM signal for comparison. In s-SNOM signal, three major processes dominate the generation of signal. The edge (Nb) launch-tip-scatter and tip-launch-edge (Nb)-scatter are reciprocal processes. They generate the $\lambda = \lambda_p$ fringes. The tip-launch-edge (hBN) reflect-tip-scatter process generate the $\lambda = \lambda_p/2$ fringes near hBN edges due to the doubled traveling path of polariton waves.

In summary, the BOSON signal arises from interference between the tip-launched polaritons and far-field illumination, with the Nb edges acting as sensitive thermal absorbers. The detected photocurrent pattern replicates the intrinsic polariton wavelength $\lambda_p$, and resembles the dominant $\lambda_p$ fringe patterns observed in s-SNOM, where the Nb edge serves as a polariton launcher and the tip acts as the detector.